%% file: paper.tex
\documentclass[11pt,a4paper]{article}
\pdfoutput=1

\usepackage{amssymb}
\usepackage[francais,english]{babel}
\usepackage{graphicx}
\usepackage{graphics}
\usepackage{lineno}
\usepackage{cite}
\usepackage[latin1]{inputenc}
\usepackage[table]{xcolor}
\usepackage{float}
\usepackage{ccaption}
\usepackage{array}
\usepackage{lscape}
\usepackage[hmargin=2cm,vmargin=3.5cm]{geometry}

\definecolor{light-gray}{gray}{0.8}

\begin{document}

\baselineskip 0.8cm

\title{Landscape genomic tests for associations between loci and environmental gradients}

\date{}

\author{Eric Frichot$^{1}$ \and Sean Schoville$^{1}$ \and Guillaume Bouchard$^{2}$ \and Olivier Fran\c cois$^{1}$}

\maketitle
\thispagestyle{empty}
\pagestyle{headings}
\setcounter{page}{1}
\pagenumbering{roman}

\begin{center}
(1) Universit\'e Joseph Fourier Grenoble, Centre National de la Recherche Scientifique, TIMC-IMAG UMR 5525, 38042 Grenoble, France.\\
(2) Xerox Research Center Europe, F38240 Meylan, France.
\end{center}

\begin{abstract}
Adaptation to local environments often occurs through natural selection acting on a large number of alleles, each having a weak phenotypic effect. One way to detect these alleles is to identify genetic polymorphisms that exhibit high correlation with environmental variables used as proxies for ecological pressures.
Here we propose an integrated framework based on population genetics, ecological modeling and statistical learning techniques to screen genomes for signatures of local adaptation. These new algorithms introduce latent factor mixed models to population genetics, employing an approach based on probabilistic principal component analysis in which population structure is introduced via unobserved variables.  These fast, computationally efficient algorithms detect correlations between environmental and genetic variation while simultaneously inferring background levels of population structure.
Comparing these new algorithms with related methods provides evidence that latent factor models can efficiently estimate random effects due to population history and isolation-by-distance patterns when computing gene-environment correlations, and decrease the number of false-positive associations in genome scans. We then apply these models to plant and human genetic data, identifying several genes with functions related to development that exhibit unusual correlations with climatic gradients.
\end{abstract}

\pagestyle{empty}

\newpage

\input{intro.tex}

\input{method.tex}

\input{results.tex}

\input{discussion.tex}

\bibliography{biblio}
\bibliographystyle{unsrt}

\input{suppmat2.tex}

\input{tables.tex}

\input{suppmat1.tex}

\end{document}

%% file: intro.tex
\section{Introduction}

Local adaptation through natural selection plays a central role in shaping the variation of natural populations \cite{Darwin_1859,Williams_1966} and is of fundamental importance in evolutionary, conservation, and global-change biology \cite{Joost_2007, Manel_2010, Barrett_2011, Schoville_2012, Jay_2012}. The intensity of natural selection commonly varies in space, and can result in gene-environment interactions that have measurable effects on fitness (e.g. \cite{Storz_2010}). Adaptive divergence can then cause local populations to evolve traits that provide advantage under local environmental conditions.

In principle, identifying chromosomal regions involved in adaptive divergence can be achieved by scanning genome-wide patterns of DNA polymorphism \cite{Nielsen_2005, Storz_2005}. Usually, the aim of screening procedures is to detect locus specific signatures of positive selection. In populations inhabiting spatially distinct environments, loci that underlie adaptive divergence can be detected by comparing relative levels of differentiation among large samples of unlinked markers \cite{Beaumont_1996, Beaumont_2004} and by using empirical tests to compare levels of differentiation to the genomic background \cite{Akey_2009, Novembre_2009}.

An alternative way to investigate signatures of local adaptation, especially when beneficial alleles have weak phenotypic effects, is by identifying polymorphisms that exhibit high correlation with environmental variables \cite{Joost_2007, Hancock2008, Poncet_2010, Pritchard_2010, Coop_2010}. In natural populations, quantitative traits that exhibit continuous geographic variation are often associated with specific variables reflecting selective pressures acting on individual phenotypes \cite{Endler_1977} . This type of variation is then reflected in geographic clines or in sympatric populations that exploit different ecological niches \cite{Haldane_1948}. Evidence for local adaptation to continuous environments can be detected if there is highly significant association with the environmental variables at some loci compared to the background genomic variation. 

However the geographical basis of both environmental and genetic variation can confound interpretation of the tests \cite{Eckert_2010}, as local adaptation can be hindered by gene flow \cite{Lenormand_2002}, and can be difficult to distinguish from the effects of genetic drift and demographic history \cite{Novembre_2009}. As a consequence, when no corrections for the effects of population structure or isolation-by-distance are considered, tests for associations between loci and environmental variables using classical regression models are prone to high rates of false-positives \cite{Meirmans_2012}. Recent studies have used the background patterns of allele frequencies to build a null model which accounts for the effects of drift and demographic history \cite{Hancock2008, Coop_2010, Hancock_2011, Fumagalli_2011}. To correct for population stratification, \cite{Hancock2008} used an empirical approach that estimates the covariance of allele frequencies among populations. These authors assessed the evidence for local adaptation of each allele by testing whether or not environmental variables explained more variance than a null model with this particular covariance structure.

Here we argue that, unless a list of a priori selectively neutral loci are used to build the empirical covariance matrix, empirical tests may face a problem of circularity. The need to identify neutral loci from the genomic background before testing implies that these tests lack power to reject neutrality. In this study, we address this problem by introducing new statistical models, called latent factor mixed models.  Using these models, we test correlations between environmental and genetic variation while estimating the effects of hidden factors that represent background residual levels of population structure. To perform parameter estimation, we extend probabilistic principal component analysis and recent statistical learning approaches \cite{Tipping_1999, Salakhutdinov_2008, Engelhardt_2010}. Based on low rank approximation of the residual covariance matrix, we implement algorithms to deal with hundreds of thousands of polymorphisms with very rapid computing times. We show that our algorithms control for random effects due to population history and spatial autocorrelation when estimating gene-environment association, and we provide examples of how our approach can be used to detect local adaptation in plants and humans.

%% file: method.tex
\section{Method}

Consider the data matrix, $(G_{i\ell})$, where each entry records the allele frequency in population or individual $i$ at the genomic locus $\ell$, $1 \leq i \leq n$, $1 \leq \ell \leq L$, and $n$ and $L$ represent the total sample size and number of loci, respectively. For simplicity, we assume our loci are biallelic, e.g. single nucleotide polymorphisms (SNPs), and data are avaiable for each individual. In this case, for each marker, there is an ancestral and a derived allele, and  $G_{i\ell}$ is the number of derived alleles for locus $\ell$ and individual $i$. For diploid data,  $G_{i\ell}$ is thus equal to 0, 1 or 2, and corresponds to the genotype at locus $\ell$. In addition to the genotypic data, we have a vector of $d$ geographic and environmental variables, $(X_i)$, for each individual. The vector of covariates could include latitude and longitude, habitat and other ecological information, climatic variables, etc, that serve as proxies for unknown environmental pressures (For example, see \cite{Hancock2008, Eckert_2010}).

{\it Model.}  To evaluate associations between allele frequencies and environmental variables while correcting for background levels of population structure, we regard the matrix $G$ as being a response variable in a regression mixed model
\begin{equation}
G_{i\ell} = \mu_{\ell} + \beta^T_{\ell} X_i + U_i^T V_{\ell} +\epsilon_{i\ell} \, ,
\label{eqn:1}
\end{equation}
where $\mu_{\ell}$ is a locus specific effect, $\beta_{\ell}$ is a $d$-dimensional vector of regression coefficients, $U_i$ and $V_{\ell}$ are scalar vectors with $K$ dimensions ($1 \leq K \leq n$). The residuals $\epsilon_{i\ell}$ are statistically independent Gaussian variables of mean zero and variance $\sigma^2$. We use Bayesian analysis to estimate the regression coefficients and their standard deviations.  We assume Gaussian prior distributions on $\mu_{\ell}$ and $\beta_{\ell j}$ with means equal to zero and variances $\sigma_{\mu}^2$ and $\sigma_{\beta_j}^2$ ($\beta_{\ell j} \sim {\rm N}(0, \sigma_{\beta_j}^2)$). Prior distributions on $U_i$ and $V_{\ell}$ are Gaussian distributions with means equal to zero and constant variance for each component (the components are independent random variables). The variance of $V_\ell$ is set to $\sigma_V^2 = 1$, and all other prior distributions on variances are non-informative. We refer to the above statistical model as a {\it Latent Factor Mixed Model} (LFMM). 
 
In LFMMs, environmental variables are introduced as fixed effects while population structure is modeled via latent factors. To separate neutral variation from adaptive variation, the matrix term $U^TV$ models the part of genetic variation that cannot be explained by the environmental pressures. Note that the use of factorization methods is closely related to estimating population structure via singular value decomposition, a well-established technique for identifying scores and loadings in principal component analysis (PCA, \cite{Jolliffe_1986}). Recently, matrix factorization methods have been generalized to include probabilistic PCA \cite{Tipping_1999} and probabilistic matrix factorization algorithms \cite{Salakhutdinov_2008}, which have proven useful in analyzing population genetic data \cite{Engelhardt_2010}. To clarify the connection between LFMM and PCA, assume that no environmental variable is available.  In this case, we set $\beta_{\ell} = 0$ for all locus $\ell$. In matrix factorization algorithms, a data matrix $G$ with $n$ rows and $L$ columns can be decomposed into a product of two matrices $U$ and $V$, where $U$ has $n$ rows and $K$ columns, and $V$ is a $K$-by-$L$ matrix. Following \cite{Patterson_2006}, we assume that the genotypic data are centered. We consider the matrix $Y_{i\ell} = G_{i\ell} - \bar{G}_{.\ell}$, where we have substracted the mean value of each column, $\bar{G}_{.\ell}  = \sum_{i = 1}^n G_{i\ell}/n$. For each individual $i$ and locus $\ell$, the decomposition is 
\begin{equation}
 Y_{i \ell}  = U_{i}^T  V_{\ell}  = \sum_{k=1}^K  U_{ik} V_{k\ell}  \, .
\end{equation}
To estimate the factor vectors $U_{i}$ and $V_{\ell}$, the squared error is minimized on the set of observed data 
\begin{equation}
\min_{U,V}   \sum_{k=1}^K   \left( Y_{i \ell}  -  U_{ik} V_{k\ell} \right)^2 \, .
\end{equation}
With $K = L$, this approach is similar to computing PCA loadings and scores \cite{Jolliffe_1986}. The number of components $K$ can, however, be chosen much lower than the number of loci or individuals. In simulations, we based our choice of $K$ on Tracy-Widom theory \cite{Patterson_2006}. In real applications, this choice of $K$ may be replaced by other estimates of population genetic structure. When values are lower than 50, our algorithm is essentially a low-rank approximation of the covariance structure \cite{Eckart_1936}, which leads to computationally fast estimation algorithms.

To simultaneously estimate scores and loadings, environmmental effects and biases, we implemented a Gibbs sampler algorithm for LFMMs (File S1). The Gibbs sampler is based on computing products of matrices of low dimension (typically, $K \leq 50$), and its speed scales with the current size of SNP data sets, around $n \approx 1,000$ and $L \approx 500,000$. In addition, we implemented a stochastic algorithm to compute standard deviations and $|z|$-scores for the environmental effects. Using the empirical distribution of $|z|$-scores obtained from all $L$ loci, we compared each locus to the genomic background and retained loci with $|z|$-scores exhibiting the highest absolute values. From a preliminary set of experiments using data simulated from the model defined in equation (\ref{eqn:1}), we found that the estimates of fixed effects stabilized quickly, after 1,000 to 10,000 sweeps for $n = 100 - 1,000$ individuals and $L = 1,000 - 100,000$ loci. A 10-fold increase in the number of sweeps, however, was necessary to recover the true values of the latent factors. Additionally, we developed numerical optimization methods to compute \textit{maximum a posteriori}  (MAP) estimates for the LFMM. One of these methods, the alternate least square method uses deterministic steps that are similar to our stochastic Gibbs sampler \cite{Bell_2007}. When checking for convergence of the MCMC algorithm, we also found that least square estimates of regression coefficients were close to the point estimates computed by the Gibbs sampler method.

{\it Theoretical considerations.}  Incorporating population genetic structure via estimates of admixture proportions or principal component analysis is common in genome-wide association studies \cite{Price_2006, Yu_2006}). \cite{Coop_2010} developed an alternative approach to identify loci underlying local adaptation in the computer program {\tt Bayenv}. To explain the difference between our approach and {\tt Bayenv}, suppose that we start by computing PCA scores from the matrix $Y$ for all individuals, and denote by $\tilde{U}_i$ the PCA scores for individual $i$. The product matrix $\tilde{U}\tilde{U}^T$ is thus equal to the empirical covariance matrix
\begin{equation}
\tilde{U}\tilde{U}^T = YY^T\!\! /n \, .  
\end{equation}
Now using the scores as covariates in a Bayesian regression model, we obtain 
\begin{equation}
G = \mu + \beta^TX + \tilde{U}^TV + \epsilon . 
\end{equation}
By a change of variables, this is equivalent to fitting the model
\begin{equation}
G =  \mu + \beta^TX + \tilde{\epsilon}
\label{eqn:bayenv}
\end{equation}
where the distribution of $\tilde{\epsilon}$ is a multivariate Gaussian distribution of the covariance matrix equal to $\sigma^2{\rm Id}+\sigma^2_V YY^T/n$ (Id is the $n$-dimensional identity matrix). Setting $\sigma^2_V = 1$ and considering small values of the scaling parameter $\sigma^2$, the model defined in equation (\ref{eqn:bayenv}) is nearly equivalent to the model implemented in {\tt Bayenv}. In a Bayesian Gaussian regression framework, incorporating PCA scores as covariates in an association model is equivalent to modeling residuals as Gaussian vectors with covariance depending on the empirical covariance matrix of the genotypic data.  Though the {\tt Bayenv} model uses a different link function, its residual term has the same covariance matrix as the genotypic data. From a theoretical point of view, the main difference between the {\tt Bayenv} model and LFMM is the inference of the factor matrix $U$ which is done in a fully Bayesian algorithm in LFMMs and in an empirical Bayes algorithm in {\tt Bayenv} (see Discussion).

%% file: results.tex
\section{Simulation Study}

We designed experiments based on simulated data to answer the following questions: 
1) Are tests based on LFMMs conservative or liberal? 
2) How does the LFMM algorithm perform compared to existing methods such as logistic or standard regression 
models \cite{Joost_2007}, principal components regression and other existing mixed models \cite{Coop_2010}? 

\paragraph{Distribution of $P$-values under the null hypothesis.}
We used equation (\ref{eqn:1}) with $\beta=0$ to generate data under a null hypothesis of no association with any environmental variables. 
In these experiments, we set the number of individuals to $n = 100$, and the number of loci to $L = 1,000$. 
We used 6 values,  $K = 1,3,5,7,10$ and $20$, for the rank of the factor matrix, $V$.  For each series of experiments, we generated 10 
replicates of this generative model, and we studied the distributions of $P$-values for tests using LFMMs.
In these tests, we set the rank of the factor matrix equal to the values we used to generate simulations. 
\begin{figure}[H]
\begin{center}
\includegraphics[scale=0.5]{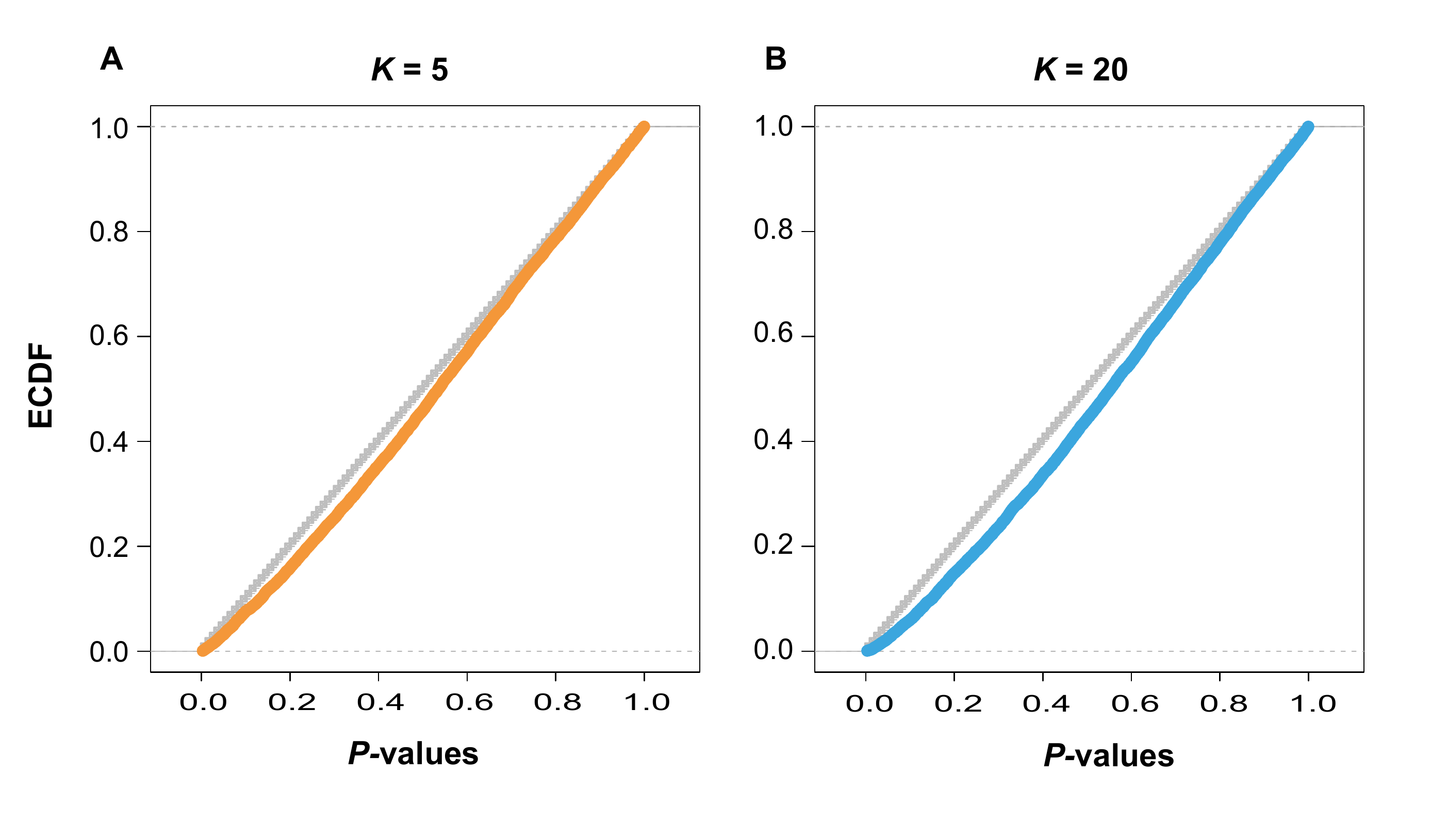}
\caption{Empirical cumulative distribution function (ECDF) for LFMM tests for simulations from a generative model using A) $K$=5, B) $K$=20 latent factors.}
\end{center}
\end{figure}

Figure 1 reports the empirical cumulative distribution function (ecdf) for $K=5$ and $K=20$. 
Plots for the other values of $K$ are shown in Figure S1. For values of $K$ lower than 5, the ecdf was close to
a uniform distribution.  For $K=20$, the tests were slightly conservative. Thus, for moderate and for large values of 
the number of latent factors, the tests produced small numbers of false positive associations.  

Next we used equation (\ref{eqn:1}) to generate data showing various levels of population structure and association with an environmental variable. 
The environmental variable was uniformly generated in the range $(0,1)$.  Here we used 3 values for the rank of the factor matrix, 
$K = 2, 20$ and $100$, representing low, moderate and high levels of underlying population genetic structure.  For each series of experiments, 
we generated 20 replicates of the generative model and compared the distribution of statistical errors for three estimation approaches: 1) LFMM, 2) standard linear regression model, 3) PC regression model. With the notations from section 2, these models were defined as follows. The LFMM was defined by equation 
$$
G_{i\ell} = \mu_{\ell} + \beta^T_{\ell} X_i + U_i^T V_{\ell} +\epsilon_{i\ell} 
$$ 
where we set the rank of the factor matrices equal to the values we used to generate simulations. The standard regression model was defined as
\begin{equation}
G_{i\ell} = \mu_{\ell} + \beta^T_{\ell} X_i +\epsilon_{i\ell} \, .
\end{equation} 
The PC regression model was defined as 
\begin{equation}
G_{i\ell} = \mu_{\ell} + \beta^T_{\ell} X_i + \tilde{U}_i^T V_{\ell} +\epsilon_{i\ell} \, ,
\end{equation}
where $(\tilde{U}_i)$ are the first $K$ PCs computed from the matrix $G$. To compute point estimates of environmental effects and their $|z|$-scores, Gibbs sampler algorithms were run for 1,000 sweeps after a burnin period of 100 sweeps. For these particular run length parameters, we checked that similar estimates were obtained for distinct initializations of the algorithm. For each locus, we recorded both the true, $B_{\ell}$, and estimated environmental effects, $\hat{B}_{\ell}$, and evaluated the absolute error 
$$
E_{\ell} = \left|  \beta_{\ell} - \hat{\beta}_{\ell}  \right| \, . 
$$

\begin{figure}[H]
\begin{center}
\includegraphics[scale=0.9]{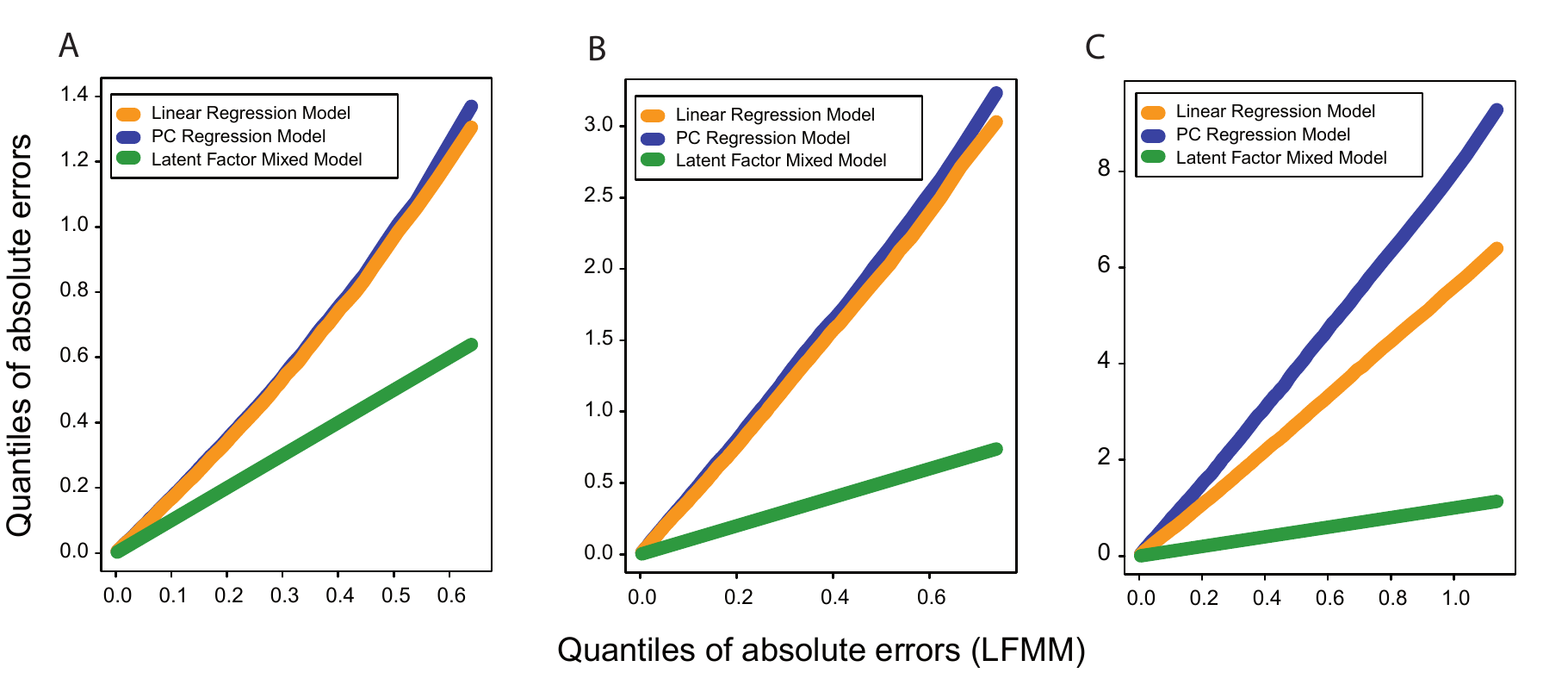}
\caption{Quantiles of absolute errors for the standard linear
regression, PC regression and LFM models using simulations from the LFM model with A) $K$=2, B) $K$=20 and C) $K$=100 latent factors.}
\end{center}
\end{figure}

Figure 2 reports the quantiles of absolute errors for the LFMM, the standard linear regression and PC regression models. 
For the LFMM, absolute errors ranged between 0 and 0.6 for $K = 2-20$, and between 0 and 1.0 for $K = 100$. 
Mean squared errors indicated that the bias and variance of estimates were small (Table 1). Compared to LFMMs, the relative errors of 
the linear and PC regression estimates increased with the rank of the hidden factor matrix.  The absolute errors of these algorithms ranged between 0 and 1.4 for $K=2$, between 0 and 3.2 for $K = 20$, and between 0 and 9.2 for $K=100$. When linear or PC regression models were fitted to the data, the quantiles of errors shifted to values $\approx$ $1.74$-fold higher for $K=2$, $\approx$ 3.8 to 4.1-fold higher for $K = 20$, and $\approx$ 5.5 to 7.7-fold higher for $K=100$. Mean squared errors provide additional evidence of relatively poor performances of the linear regression and PC regression estimates when levels of underlying structure increase (Table 1).   

\paragraph{Spatial coalescent simulations.}

In another series of experiments, we compared the LFMM estimation algorithm against two methods that do not correct for population stratification, 
and against two methods that use the empirical covariance matrix as a correction. The first set of methods include a linear model and generalized linear model 
(LM and GLM, \cite{Joost_2007}), and the second set of methods include a PC regression model (PCRM) and the mixed model 
{\tt Bayenv} \cite{Coop_2010}. To enable comparisons, we simulated genotypic data from spatial coalescent models with the computer program {\tt ms} 
\cite{Hudson_2002}. Ten data sets were generated according to a linear stepping-stone model with 40 demes, setting the effective migration 
rate between pairs of adjacent demes to the value $4Nm = 25$. Sampling 5 individuals in each deme, each data set included a total of $n=200$ 
haploid individuals genotyped at $L = 1,000$ SNP loci. Using Tracy-Widom tests implemented in {\tt SmartPCA}, we found that the number of principal 
components with $P$-values smaller than $0.01$ was around $K_{\rm TW} = 7$. We ran the Gibbs sampler algorithm during 100 sweeps for burnin, and we used the 
next 900 sweeps to compute points estimates, variances and $|z|$-scores.

\paragraph{Distribution of $P$-values.} To examine the outcome of tests when genetic variation is neutral at all loci,  
we computed the distributions of $P$-values under a LM, GLM, PCRM and LFMM with different values for the number of latent factors
($K$ ranging from 1 to 20). The distributions of $P$-values for tests based on LM and GLM showed a strong departure from the 
uniform distribution (Figure 3A-B). In those cases, the tests were too liberal, and produced a large number of false positive 
results. Using $K = 7$ latent factors or PCs, the distribution of $P$-values for tests based on an LFMM or PCRM was much closer to a uniform 
distribution (Figure 3C-D). In addition, we found that choosing $K$ based on Tracy-Widom theory led to slightly conservative tests for the 
particular simulation settings used here. Ecdf for all values of $K$ are shown in Figures S2 and S3, respectively.

\begin{figure}[H]
\begin{center}
\includegraphics[scale=0.5]{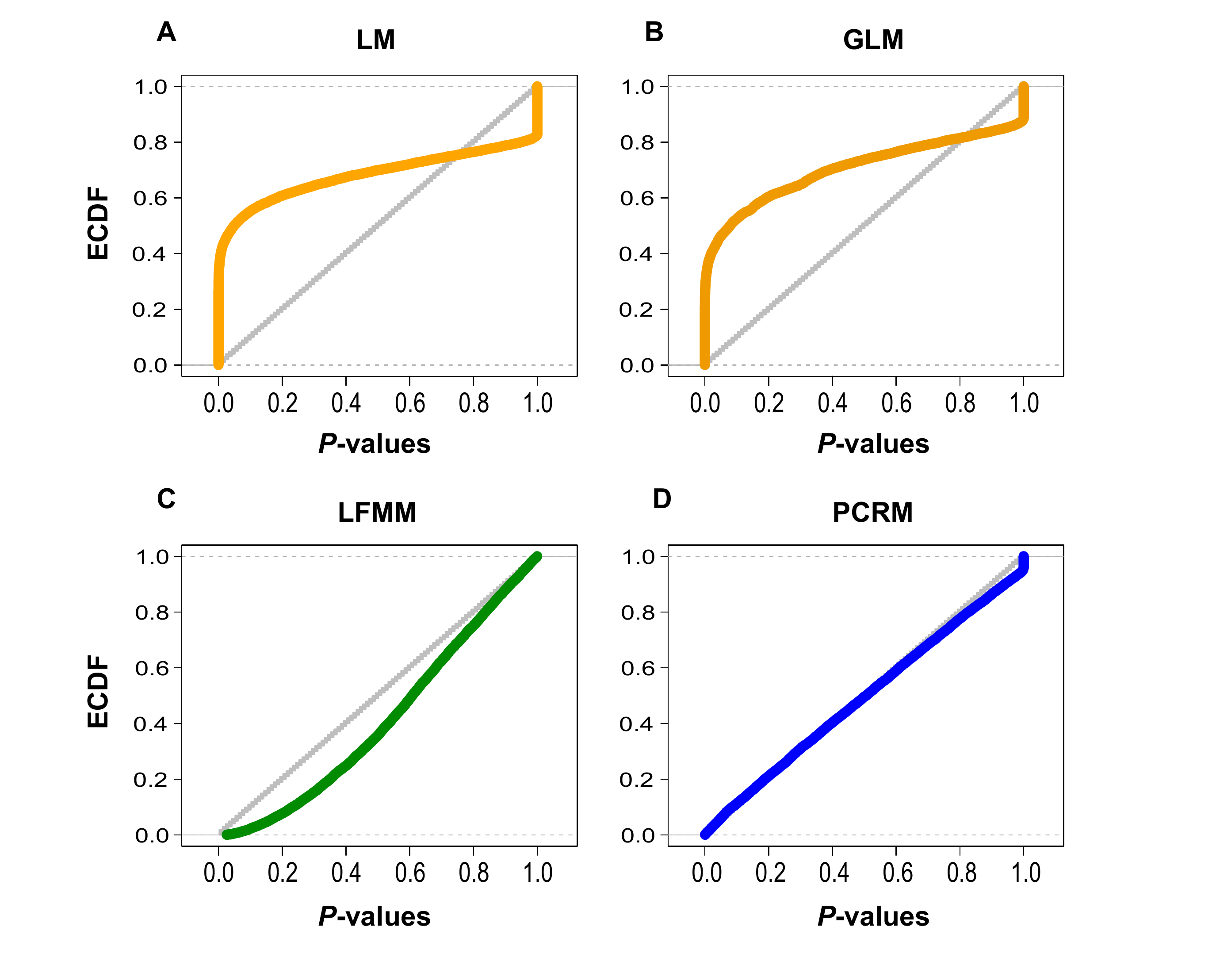}
\caption{Empirical cumulative distribution function (ECDF) for A) the linear regression model (LM), B) the generalized linear model (GLM), C) the LFM model using $K=7$ latent factors (LFMM), and D) the PC regression model using $K=7$ principal components (PCRM).}
\end{center}
\end{figure}

Next we evaluated the ability of LFMMs to detect loci exhibiting correlations with particular environmental gradients, and compared tests based on 
an LFMM with methods based on linear models and the computer program {\tt Bayenv} \cite{Coop_2010}. 
An environmental variable, $x$, was defined for each population as the geographic identifier of the population in the linear stepping-stone
model. We created an environmental gradient using a logistic function, $s(x)$, of $x$ as follows 
\begin{equation}
s(x) = \frac{1}{1+e^{\theta(x-20)}} \, , \quad \theta > 0.
\end{equation}
For each of the 10 previously generated neutral stepping-stone simulations, we simulated binary alleles for each deme $x$ at 50 
loci with frequency $s(x)$, with the slope of the gradient $\theta = 0.2$.
We then obtained 10 data sets with $L = 1050$ loci including 50 loci correlated with the environmental gradient, $s(x)$.
Across these datasets, {\tt smartPCA} estimates between 5 and 7 significant eigenvectors ($P<0.01$ in the Tracy-Widom test). For the simulated data sets, we evaluated the percentage of false negative (FN) and of false positive (FP)
 tests based on LM, GLM, PCRM and LFMM, for two values of the type I error (Table 2).

Using $\theta = 0.2$ in simulations of non-neutral loci, we found that linear models exhibited high percentages of FP. 
In contrast, tests based on PCRMs exhibited very large percentages of FN, and had no power to reject neutrality. 
Tests based on LFMM produced low numbers of FP, and had reasonable power to reject the null hypothesis of no association.

\paragraph{Comparisons with {\tt Bayenv}.} 
For each of the 10 previously generated neutral stepping-stone simulations, we simulated binary alleles for each deme $x$ at 50 loci with frequency $s(x)$, using $\theta = 0.1$.
To enable comparision with the program {\tt Bayenv}, which returns Bayes factors instead of $P$-values, we considered ranked lists 
recording the $M$ loci corresponding to the strongest (true) associations ($M$ between 1 and 1,050). For each $M$,
we computed the number of true positives and the number of false negatives. Locus ranking
 was performed on the basis of $|z|$-scores in LFMM, and on the basis of Bayes factors in {\tt Bayenv}. 
The LFMM tests used values of $K$ equal to $K = 1,3,5,7,10$ and $20$, and we used the default parameters 
of the {\tt Bayenv} algorithm  to compute Bayes factors (run length of 30,000 sweeps). Experiments were assessed by measuring the area under the receiver-operating characteristic curve (AUC) averaged over 10 replicates.
The mean AUC for tests based on LFMM with $K=5-7$ factors were around $0.95-0.96$ whereas the AUC for {\it Bayenv} was equal to 0.88. 
In the linear stepping stone model simulations, the tests based on LFMM obtained better performances than {\tt Bayenv} 
for all values of $K$ (Figure 4).

\begin{figure}[H]
\begin{center}
\includegraphics[scale=0.4]{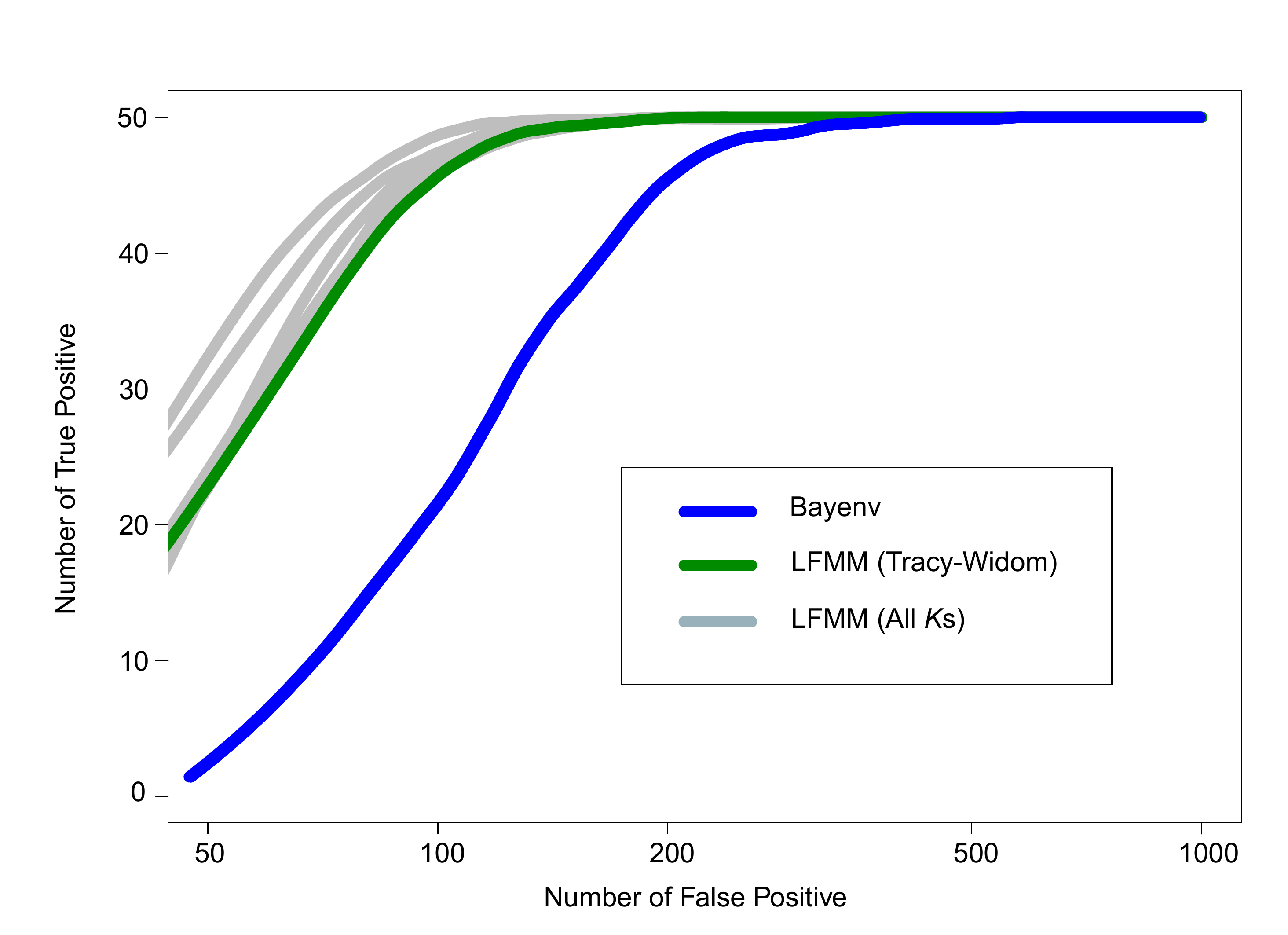}
\caption{Detection of outlier loci. Number of true positives for the {\tt Bayenv} model and the Latent Factor Mixed Model for $K=7$ (Tracy-Widom value) and for $K=1,3,5,10,20$ for spatial coalescent simulations including loci under selection.}
\end{center}
\end{figure}

\section{Data Analysis}

To illustrate the application of LFMMs, we analyzed genomic data sets of Loblolly Pines ({\it Pinus taeda}, Pinaceae, \cite{Eckert_2010}) 
and humans (Human Genome Diversity Panel, \cite{Li_2008}).

\subsection*{Loblolly Pine}

The Loblolly Pine is distributed throughout the southeastern USA, ranging from the arid Great Plains to the humid Eastern Temperate Forest ecoregion. The data consisted of 1,730 SNPs selected in expressed sequence tags (EST) for 682 individuals \cite{Eckert_2010}. Following \cite{Eckert_2010}, we considered 5 environmental variables representing the five first components of a PCA for 60 climatic variables. The first component (PC1) was mainly described by latitude, longitude, temperature and winter aridity. PC2 was described by longitude, spring-fall aridity and precipitation \cite{Eckert_2010}. 

For each of the 5 environmental variables, we applied the LFMM algorithm using 100 sweeps for burnin and 400 additional sweeps to 
compute $|z|$-scores for all loci. Based on a prior analysis of the genotypic data with the program {\tt SmartPCA}, we used $K=10$ latent factors. 
A total of 392, 113 and 30 SNPs obtained $|z|$-scores greater than 3, 4 or 5 for at least one environmental variable, respectively. 
Based on this result, we considered that a SNP effect was significant when its $|z|$-score was greater than 4 (two-sided test). 
Among the 50 loci with the highest $|z|$-scores, 17 were shared with those detected by \cite{Eckert_2010} using {\tt Bayenv}. Seven of the 10 SNPs with Bayes factors greater than 
$10^3$ were confirmed by the LFMM analysis. For the first and second enviromnental variables, the two SNPs which obtained 
the highest Bayes factors using {\tt Bayenv} were recovered by the LFMM analysis. Table 3 provides a list of SNPs associated with climatic gradients 
and their functional annotation. 
Compared to the analysis of \cite{Eckert_2010}, the LFMM analysis discovered new significant and interesting associations with climatic gradients, 
for example, the chloroplast lumen 19 kDA protein involved in photosynthesis ($|z| = 6.42$), a pentatricopeptide repeat protein involved in 
oxidative stress and salt stress ($|z| = 5.90$), and the heat shock transcription factor hsf5 ($|z| = 5.60$) involved in regulation of 
transcription and response to temperature stress (Table 3 and Table S1).

\subsection*{Human data analysis}

We applied an LFMM analysis to a worldwide sample of genomic DNA from 1,043 individuals in 52 populations, referred to as the Human Genome 
Diversity Project -- Centre Etude Polymorphism Humain (HGDP-CEPH) Human Genome Diversity Cell Line Panel (hagsc.org/hgdp/). The genotypes were 
generated on Illumina 650K arrays \cite{Li_2008}, and the data were filtered to remove low quality SNPs included in the original files.

We extracted climatic data for each of the 52 population samples using the WorldClim data set at 30 arcsecond ($1 { \rm km}^2$) resolution 
\cite{Hijmans_2005}. These data include 11 bioclimatic variables interpolated from global weather station data collected during a 50 year period 
(1950-2000). The climatic variables included annual mean temperature, mean diurnal range, maximum temperature of warmest month, minimum temperature 
of coldest month, annual precipitation, etc (Table S2). We summarized the climatic variables by using the first axis of a principal component 
analysis. For this first principal component, we applied the LFMM algorithm to compute $|z|$-scores for each locus with $K=50$ latent factors, 
using 100 sweeps for burnin and 900 additional sweeps to warrant convergence. 

A total of 2,624 (0.4\%), 508 (0.08\%) and 65 (0.007\%) SNPs obtained $|z|$-scores greater than 5, 6 or 7, respectively 
(Figure S3). Among loci with $|z|$-scores greater than 5, 28 GWAS-SNPs with known disease or trait association were found \cite{Hindorff_2009}). 
These include several SNPs discovered by \cite{Hancock_2011}. For example the SNPs rs12913832 and rs28777, $|z|$-scores greater than 6, are associated with genes OCA2 and SLC45A2 (Table 4). Among the SNPs significantly correlated with climatic gradients, several notable examples include genes 
associated with celiac disease (ICOSLG),  height (LHX3-QSOX2 and IGF1), and vitamin D synthesis or activation (NADSYN1 encoding nicotinamide 
adenine dinucleotide synthetase and DHCR7 the gene encoding 7-dehydrocholesterol reductase, an enzyme catalyzing the production in skin of 
cholesterol from 7-dehydrocholesterol (Table 4).
Among the 65 SNPs with $|z|$-scores greater than 7, 31 were located within genes (Table S3).
One interesting result is that several genes are associated with multicellular organismal development. EPHB4
($|z|= 8.90$) is involved in heart morphogenesis and angiogenesis,
NRG1 ($|z|= 7.15$) involved with nervous system development and cell proliferation, RBM19
($|z| = 7.04$) involved with positive regulation of embryonic development, EYA2 ($|z|= 7.09$)
involved with eye development and DNA repair, and POLA1 ($|z| = 7.63$) involved with the mitotic cell cycle
and cell proliferation \cite{Saccone_2011, Hornbeck_2012}.

%% file: discussion.tex
\section{Discussion}

\paragraph{Interpretation of LFMM results and other methods.} Based on a matrix factorization approach, LFMMs incorporate a unified framework for estimating effects of environmental and demographic factors on genetic variation. Without environmental variables, LFMMs are equivalent to performing a probabilistic PCA of allele frequencies \cite{Tipping_1999}. When environmental variables are included, hidden factors capture the part of genetic variation that cannot be explained by the set of measured environmental variables. This fraction of genetic variation could result from the demographic history of the species, unknown environmental pressures or from IBD patterns.

While a plethora of statistical tests have been proposed for detecting genes evolving under positive selection and local adaptation \cite{Storz_2005, Novembre_2009}, the development of tests based on correlations with habitat or landscape variables is still recent \cite{Joost_2007, Hancock2008}. Compared to methods based on summary statistics, tests based on environmental association have increased power to detect selection from standing genetic variation and soft sweeps in a species genome \cite{Pritchard_2010, Schoville_2012}. However, simple implementation of these tests, for example simple linear or logistic regression models, can be misleading in the presence of IBD patterns \cite{Meirmans_2012}. Our simulation results provide clear evidence that tests based on LFMMs significantly reduce the rates of false positive associations in the presence of IBD.  

While both the mixed model approach of the computer program {\tt Bayenv} and the LFMM approach includes a covariance structure in a regression model, there are important differences between the two approaches. A first improvement is that LFMMs estimate latent factors and regression coefficients simultaneously, while {\tt Bayenv} first estimates a covariance matrix, and then uses it when estimating (random) environmental effects. To apply {\tt Bayenv}, the authors suggest utilizing selectively neutral SNPs to estimate the covariance matrix. This approach requires separating neutral from adaptive variation a priori, and is difficult to apply when selection acts on phenotype at a large number of loci. Inclusion of adaptive markers in the "neutral set" is sometimes unavoidable, and in this case, {\tt Bayenv} may overlook interesting associations. This distinction between approaches explains the observed differences in the lists of outlier loci for Loblolly pines, where 1,730 SNPs were genotyped in expressed sequences. For these data it was extremely difficult to select neutral SNPs from the background a priori.  Another distinction between LFM and {\tt Bayenv} approaches is our use of low rank approximations of the covariance matrix. LFMMs actually estimate correlations between environmental predictors and allele frequencies while $K$ hidden factors explain residual genetic variation, where $K$ is much smaller than the sample size. The low rank approximation is computationally faster than {\tt Bayenv} when analyzing large data sets. 

\paragraph{Number of latent factors.} In the LFM modeling approach, the choice of low values for $K$ is important for optimizing the computational performances of the estimation algorithm. This choice is reminiscent of selecting the number of components in PCA or in Bayesian clustering programs, and it has also an impact on test outcomes.  For values of $K$ taken too large, the tests are highly conservative, and the power to reject neutrality declines. Values of $K$ that minimize the trade-off between the bias and variance for our statistical estimates could be obtained by using cross-validation procedures, but cross-validation procedures are computationally intensive, so instead we use Tracy-Widom theory to select $K$ \cite{Patterson_2006}. We evaluated this choice during our simulation analysis, and found that it led to slightly conservative tests. Although the choice of Tracy-Widom estimates is suboptimal, the performances of LFMMs were still superior to those of {\tt Bayenv} in simulations of IBD patterns. In the analysis of human data, we restricted $K$ to be less than 50 (approximately the number of population samples).  We suggest that, when there is a reasonable estimate of the number of genetic cluster for a species, it can be used in LFMM tests directly.  While finer grain population structure could be evaluated \cite{Lawson_2012}, our choice was again motivated by a trade-off between accuracy and run-time. A future development of our LFMM approach will be to develop fast numerical optimization procedures based on variational approximations of the likelihood, which will allow us to implement cross-validation algorithms and increase the power of tests.

\paragraph{Plant and human data.} For {\it Pinus taeda},  the LFMM results confirmed that several ESTs previously discovered with {\tt Bayenv} had functions linked to climate \cite{Eckert_2010}. In addition, the LFMM analysis discovered new interesting candidate SNPs. Those variants include functions associated with wound repair and immunity, photosynthetic activity and carotenoid biosynthesis, cellular respiration and carbohydrate metabolism, heat, salt and oxidative stress responses (Table 3). Applying LFMMs to the HGDP data, we found that a total of 0.4\% of all polymorphisms (2,624 SNPs) exhibited significant associations with temperature gradients ($|z| > 5$). For example, we identified SNPs associated with the gene OCA2 that may be functionally linked to blue or brown eye color and the gene SLC45A2 that may be associated with skin pigmentation \cite{Hancock_2011}. This list also contained SNPs associated with height and vitamin D synthesis and diseases such as gluten intolerance and Crohn's disease. Our list of genic SNPs with $|z|$-scores greater than 7 ($|z| > 7$) was enriched in genes involved in organismal development. For example, the genes EPHB4, BOK, and NRG1 $-$with functions related to heart and brain development$-$ were associated with climatic gradients. Overall, the analysis confirmed that many loci are associated with climatic gradients or to correlated evolutionary pressures (for example, pathogenic environment). This result supports the hypothesis that soft sweeps may have been common in recent human evolution \cite{Pritchard_2010}.

\paragraph{Conclusion.} With ever increasing amounts of genetic data generated by high-throughput sequencing technologies, population genetic methods have shifted from traditional statistical approaches to approaches that use statistical learning techniques. Estimates of ancestry and other population parameters are commonly obtained from mixture models \cite{Pritchard_2000, Durand_2009, Alexander_2011}, principal component analyses \cite{Patterson_2006}, hidden Markov models \cite{Price_2009} and factor analysis \cite{Engelhardt_2010}. Our study contributes to the machine learning toolbox for population and landscape genomic analysis by implementing new gene-environment association tests based on matrix factorization methods. 

\paragraph{Software availability.} Source codes and computer programs for fitting LFMMs are available from the authors web-sites.

\paragraph{Acknowlegments.}  This work was supported by a grant from la R\'egion Rh\^one-Alpes to Eric Frichot and Olivier Fran\c{c}ois, and by an NSF grant to Sean Schoville (OISE-0965038). Olivier Fran\c{c}ois acknowledges support from Grenoble INP.
 

%% file: suppmat2.tex
\section*{Supplementary Text 2}

\begin{figure}[H]
\legend{FIGURE S1: Empirical cumulative distribution function for LFMM tests for simu-
lations from generative models with $K=1,3,5,10,$ and $20$ latent factors.}
\begin{center}
\includegraphics[scale=0.5]{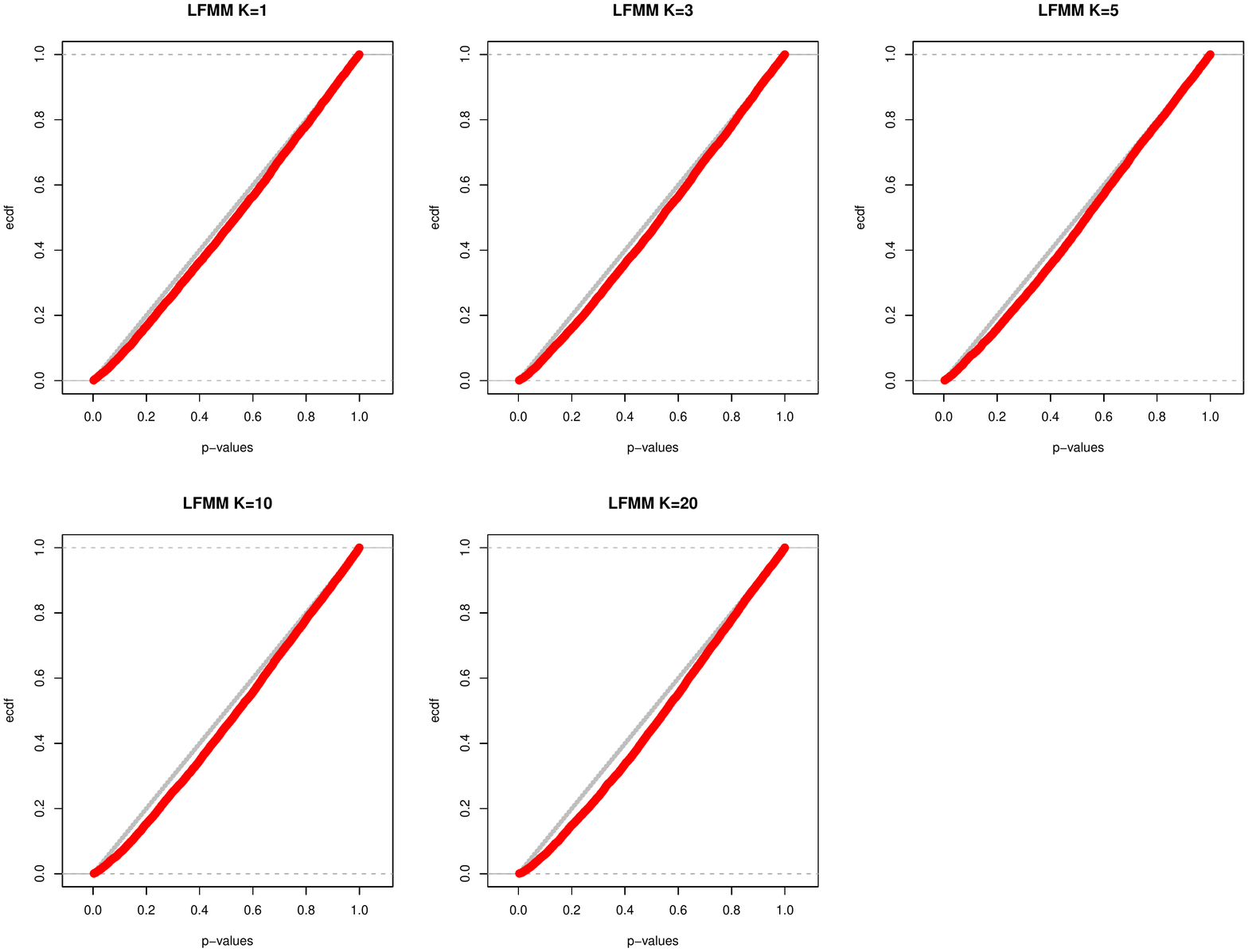}
\end{center}
\end{figure}

\begin{figure}[H]
\legend{FIGURE S2: Empirical cumulative distribution function for the LFMM using $K = 1,3,5,7,10,$ and $20$ latent factors.}
\begin{center}
\includegraphics[scale=0.8]{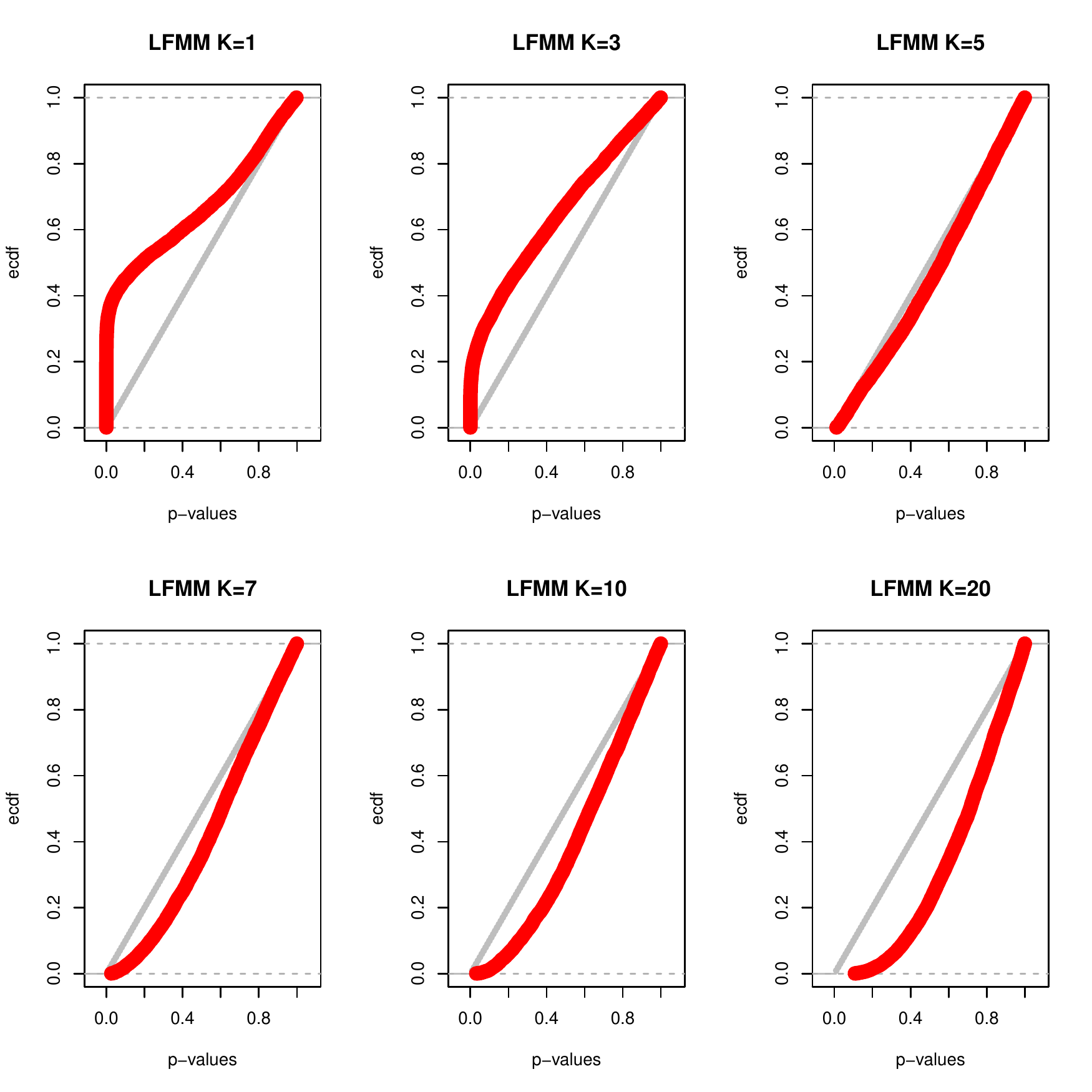}
\end{center}
\end{figure}

\begin{figure}[H]
\legend{FIGURE S3: Empirical cumulative distribution function for the PC regression model using $K = 1,3,5,7,10,$ and $20$ latent factors.}
\begin{center}
\includegraphics[scale=0.8]{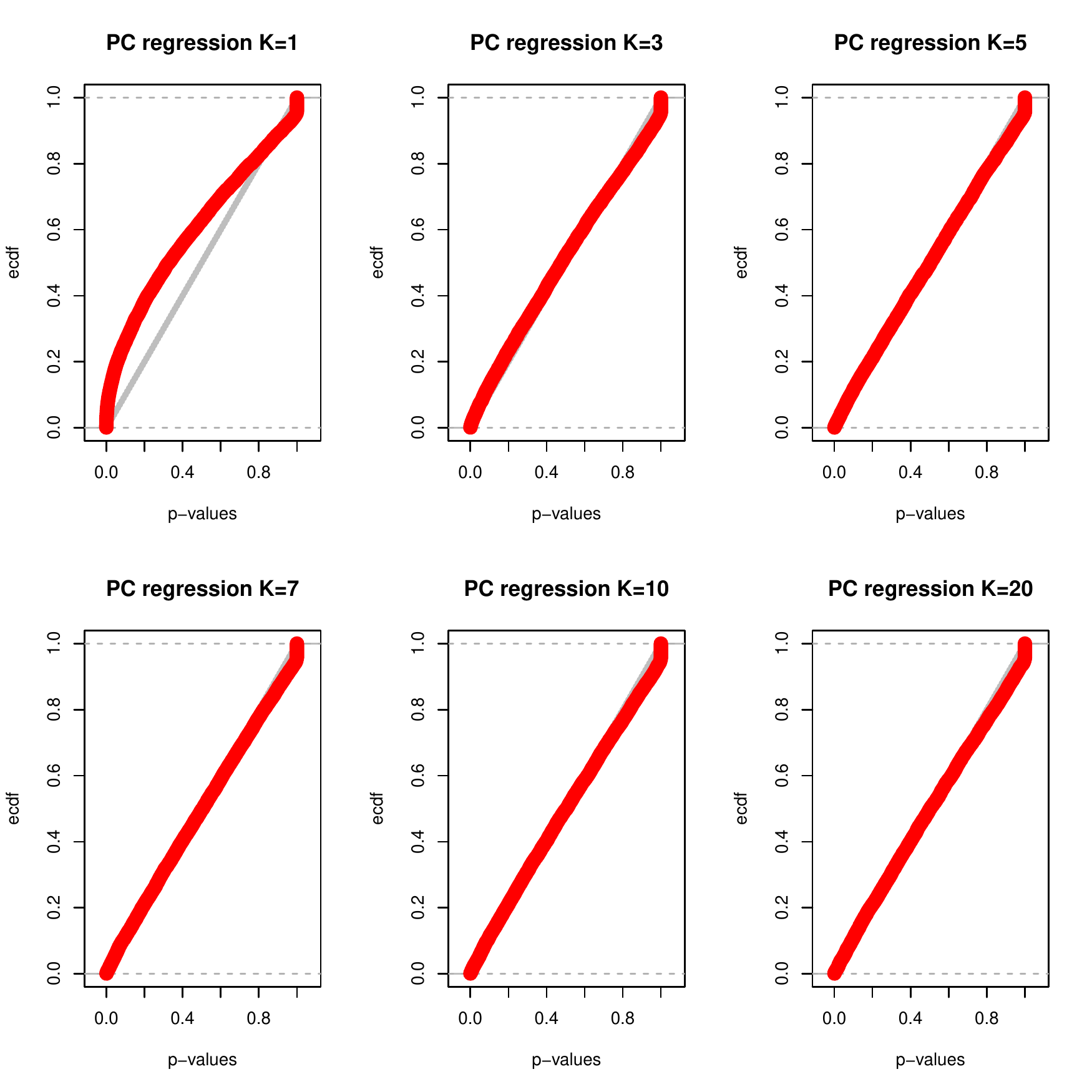}
\end{center}
\end{figure}

%% file: tables.tex
\section*{Tables}

\begin{landscape}
\begin{table}[H]
\legend{TABLE 1: Mean squared errors for estimates of environmental effects.}
\begin{center}
\begin{tabular}{cccc}
 & & & \\
$K$ &  LM & PCRM & LFMM \\
\hline
2    &    0.20 &    0.21 & 0.15\\
20   &   1.27    &  1.42 & 0.08\\
100  &   6.13   &   12.41 & 0.20\\
\end{tabular}
\end{center}
\end{table}
\newpage
\begin{table}[H]
\centering
\legend{TABLE 2: Percentage of false negative, FN and false positive (FP)
for linear, PCR and LFM models and type I error $\alpha$. }

\begin{tabular}{c|cccc}
        FN (FP) & LM & GLM & PCRM & LFMM\\
    \hline
        type I error: & & & & \\
        $-\log_{10} \alpha = 3$ & 0\% (33\%) & 0\% (24\%) & 100 \% (3\%) & 4\% (5\%) \\
        $-\log_{10} \alpha = 4$ & 0\% (27\%) & 0\% (19\%) & 100 \% (0\%) & 14\% (3\%) \\
\end{tabular}
\end{table}
\newpage
\begin{table}[H]
\legend{TABLE 3: Loblolly Pines. Annotation and gene ontology for some interesting SNPs with $z$-scores with absolute value greater than 4 for the first two components of 60 climatic variables.}
\begin{center}
\begin{tabular}{l|l|c}
Annotation 				& 		Gene Ontology 			& $-\log_{10}$($P$-value) \\
\hline 
Thylakoid lumenal 19 kda chloroplast 	& Oxygen evolving complex; Photosystem II 	& $9.87$\\
Pentatricopeptide repeat protein 	& Oxidative stress; Salt stress 		& $8.44$\\
\textbf{Conserved hypothetical protein} & Ubiquitin-specific protease 			& $8.28$\\
Chalcone synthase  			& Flavonoid biosynthesis; 
					  Wound response; Oxidative stress		& $7.80$\\
Heat shock				& Temperature stress				& $7.67$\\
Dirigent protein pdir18			& Disease response				& $6.56$\\
Heat shock transcription factor hsf5	& Regulation of transcription; 
					  Response to stress				& $6.15$\\
Zinc finger				& Transcription; DNA binding; Zinc ion binding	& $5.84$\\
Probable n-acetyltransferase hookless 1	& Auxin signaling; Photomorphogenesis; 
					  Ethylene response				& $5.78$\\
\textbf{Calcium-binding pollen allergen}& Polcalcin; Calcium ion binding		& $4.61$\\
Geranylgeranyl diphosphate synthase	& Cholesterol Biosynthesis; 
					  Isoprenoid biosynthesis			& $4.59$\\
\textbf{Hypothetical protein OsI\_04393} & Trehalose-6-phosphate phosphatase		& $4.59$\\
\hline 
\textbf{Potassium proton antiporter}	& Potassium ion transport; 
					  Solute:hydrogen antiporter			& $5.54$\\
\textbf{DNA mismatch repair}		& DNA repair; Regulation of DNA recombination	& $5.44$\\
\end{tabular}
\end{center}
\end{table}

\begin{table}[H]
\footnotesize
\legend{TABLE 4: Human data. HGDP SNPs with the highest $z$-scores among those associated with phenotypic traits in GWAS.}
\begin{center}
\begin{tabular}{p{3cm}|l|l|l|c}
\textbf{Landscape-Trait category} & \textbf{Ref SNP ID} 	&	\textbf{Nearby gene}		& 	\textbf{Disease or trait association} & $-\log_{10}$($P$-value) \\
\hline
Pigmentation& rs32579     & PPARGC1  &  Tanning 					&$9.42$\\
and tanning & rs12913832  & OCA2/HERC2         &  Eye color, Eye color traits, Hair color, 					&$9.15$\\
 &             &          &  Black vs. blond hair color, 		&\\
 &             &          &  Black vs. red hair color 			&\\
 & rs11234027	& DHCR7	  &  Vitamin D levels				&$7.78$\\
 & rs3129882   & HLA-DRA  &  Parkinson's disease 			&$6.97$\\
 & rs28777     & SLC45A2  &  Black vs. blond hair color 		&$6.90$\\
 &             &          &  Black vs. red hair color 			&\\
Immune and & rs1250550   & ZMIZ1    &  Crohn's disease 				&$8.77$\\
autoimmune &             &          &   Inflammatory bowel disease (early onset) 	&\\
 & rs2735839   & KLK3     &  Prostate cancer 				&$8.16$\\
 & rs9264942   & RPL3P2         &  HIV-1 control				&$8.02$\\
 & rs2179367   & Intergenic between SUMO4 and ZC3H12D         &  Dupuytren's disease 			&$7.57$\\
 & rs1551398   & Intergenic between TRIB1 and LOC100130231         &  Crohn's disease 				&$7.45$\\
 & rs2289700   & CTSH     &  Bipolar disorder 				&$6.98$\\
 & rs4819388   & ICOSLG     &      Celiac disease				&$6.67$\\
 & rs703842    & CYP27B1/METTL1         &  Multiple sclerosis 			&$6.59$\\
 & rs12593813  & MAP2K5   &  Restless legs syndrome 			&$6.40$\\
 & rs4664308   & PLA2R1   &  Nephropathy (idiopathic membranous) 	&$6.28$\\
Metabolism & rs10908907  & Intergenic MUC7         &  Alcoholism (heaviness of drinking)&$8.91$\\
 & rs1566039   & Intergenic between PAPD7 and MIR4278         &  Sphingolipid levels 			&$6.89$\\
 & rs7665090   & MANBA         &  Primary biliary cirrhosis 			&$6.48$\\
Cardiovascular & rs869244    &  ADRA2A        &  Platelet aggregation 			&$7.20$\\
 & rs12034383  & CR1         &  Erythrocyte sedimentation rate  		&$7.15$\\
 & rs3129882   & HLA-DRA         &  Systemic sclerosis 			&$6.97$\\
 & rs11897119  &  MEIS1        &  PR interval				&$6.71$\\
Height & rs7678436   & NCAPG-LCORL         &  Height 					&$9.43$\\
Other & rs12479254  &  BOK        &  Brain structure 				&$9.43$\\
\end{tabular}
\end{center}
\end{table}

\begin{table}[H]
\footnotesize
\legend{TABLE S1: Loblolly pines. SNP identifier and annotation for SNPs with $z$-scores with absolute value greater than 4 for the first two components of 60 climatic variables.}

\begin{center}
\begin{tabular}{l|l|c}
SNP & Annotation & $-\log_{10}$(P-value) \\
\hline
2-4107-01-438 & thylakoid lumenal 19 kda chloroplast			& $9.87$\\
0-10719-01-95 & pentatricopeptide repeat protein			& $8.44$\\
\textbf{2-1087-01-86}& conserved hypothetical protein [Ricinus communis]& $8.28$ \\
2-1818-01-168 & chalcone synthase					& $7.80$ \\
CL17Contig1-03-443 & heat shock						& $7.67$ \\
0-9449-02-292 & dirigent protein pdir18                 		& $6.56$ \\
0-18317-01-495 & potassium proton antiporter				& $6.46$ \\
UMN-CL194Contig1-04-130 & dna mismatch repair				& $6.24$ \\
\textbf{0-17238-01-294} & Nodulin MtN21 family protein			& $6.20$ \\
0-17776-01-96 & heat shock transcription factor hsf5			& $6.15$ \\
0-8823-01-306 & squamosa promoter-binding 				& $5.91$ \\
2-4856-01-162 & zinc finger						& $5.84$ \\
0-4838-01-307 & probable n-acetyltransferase hookless 1			& $5.78$ \\
2-3236-01-225 & arabinogalactan-like protein				& $5.76$ \\ 
0-768-02-400 & protein kinase family protein				& $5.72$ \\
UMN-5299-01-201 & importin-alpha re-					& $5.33$ \\
2-4724-01-136 & protein kinase						& $4.98$ \\
0-18887-02-633 & amino acid transporter					& $4.94$ \\
CL996Contig1-03-68 & af448201 1 alpha-xylosidase			&$4.93$ \\
2-3884-02-413 & sf21d1 splice variant protein				&$4.92$ \\
UMN-CL148Contig1-02-220 & Histone 2					&$4.82$ \\
CL3851Contig1-05-68 & proliferating cell nuclear antigen		&$4.81$ \\
CL2121Contig1-05-658 & glycolipid transfer				&$4.74$ \\
\textbf{CL763Contig1-06-141} & calcium-binding pollen Polcalcin		&$4.61$ \\
0-16664-01-58 & geranylgeranyl diphosphate synthase			&$4.59$ \\
\textbf{UMN-1598-02-647} & hypothetical protein OsI 04393 [Oryza sativa Indica Group]&$4.59$ \\
2-7619-01-193 & target of myb1						&$4.57$ \\
2-2125-01-274 & nodulation receptor kinase				&$4.40$ \\
CL3162Contig1-02-257 & small gtp-binding protein			&$4.32$ \\
0-13722-01-343 & dirigent-like protein					&$4.22$ \\
\hline 
\textbf{0-8922-01-655} & TIFY domain containing protein			&$6.01$ \\
\textbf{0-18317-01-495} & potassium proton antiporter			&$5.54$ \\
\textbf{UMN-CL194Contig1-04-130} & dna mismatch repair			&$5.44$ \\
CL1381Contig1-01-188 & aintegumenta-like protein			&$5.26$ \\
CL3851Contig1-05-68 & proliferating cell nuclear antigen		&$4.52$ \\
2-2125-01-274 & nodulation receptor kinase				&$4.27$ 
\end{tabular}
\end{center}
\end{table}

\begin{table}[H]
\legend{TABLE S2: Climatic variables used in the analysis of the HGDP data set.}
\begin{center}
\begin{tabular}{l|c}				
BIO1 & Annual Mean Temperature			\\
BIO2 & Mean Diurnal Range (Mean of monthly (max  - min)	\\
BIO3 & Isothermality (BIO2/BIO7)			\\
BIO4 & Temperature Seasonality (standard deviation * 100)\\
BIO5 & Max Temperature of Warmest Month			\\
BIO6 & Min Temperature of Coldest Month			\\
BIO7 & Temperature Annual Range  (BIO5-BIO6)		\\
BIO8 & Mean Temperature of Wettest Quarter		\\
BIO9 & Mean Temperature of Driest Quarter		\\
BIO10 & Mean Temperature of Warmest Quarter		\\
BIO11 & Mean Temperature of Coldest Quarter 		\\
\end{tabular}
\end{center}
\end{table}

\begin{table}[H]
\scriptsize
\legend{TABLE S3: Human data. HGDP SNPs with $z$-scores with absolute value greater than 7 in genes with molecular (Mol), and biological (Bio) functions associated with these genes.} 
\begin{center}
\begin{tabular}{c|c|c|c|c|p{12cm}}				
SNP   &          CHR  &   BP    &          Gene     &       |z|-score	&	Gene function\\
\hline
7529482   &      1    &   203659355  &     ATP2B4/intron &  7.15 & 	(Mol)   calmodulin binding; protein binding; hydrolase activity; calcium-transporting ATPase activity; metal ion binding; nucleotide binding; ATP binding; hydrolase activity, acting on acid anhydrides, catalyzing transmembrane movement of substances; PDZ domain binding\\
          &		&		&		&	&	(Bio)   platelet activation; transport; ATP biosynthetic process; blood coagulation; transmembrane transport; cation transport\\
3816186   &      2      & 42936547      &  MTA3/nearGene-3& 7.35& (Mol)  zinc ion binding; sequence-specific DNA binding; metal ion binding; transcription factor activity\\
4681618   &      3      & 150146026     &  TSC22D2/intron&  7.25 & (Mol)   transcription factor activity\\
9784335   &      3      & 150159767     &  TSC22D2/intron&  7.21 & (Bio)   response to osmotic stress \\
10935800  &      3      & 150149696     &  TSC22D2/intron&  7.36 & \\
11708779  &      3      & 55934939      &  ERC2/intron   &  7.40 & (Mol) protein binding\\

144173    &      7   &    100416250     &  EPHB4/cds-synon& 8.90& (Mol)   protein binding; protein-tyrosine kinase activity; ephrin receptor activity; nucleotide binding; transmembrane receptor protein tyrosine kinase activity; receptor activity; ATP binding \\
	&	     &			&		 &	& heart morphogenesis; cell migration during sprouting angiogenesis; protein amino acid autophosphorylation; multicellular organismal development; ephrin receptor signaling pathway; angiogenesis; cell adhesion\\
3807496   &      7   &    16821355      &  TSPAN13/intron & 7.46&  -- \\
4729616   &      7   &    100462565     &  SLC12A9/intron & 7.35&  (Mol)   cation:chloride symporter activity \\
	  &	     &			&		  &     &  (Bio)   transmembrane transport\\
6942733   &      7   &    100350763     &  ZAN/missense   & 7.41& (Bio)   cell-cell adhesion; binding of sperm to zona pellucida\\
10953303  &      7   &    100365613     &  ZAN/missense   & 7.37& \\

989465    &      8   &    32105334      &  NRG1/intron    & 7.04& (Mol)   protein binding; transmembrane receptor protein tyrosine kinase activator activity; growth factor activity; ErbB-3 class receptor binding; cytokine activity; transcription cofactor activity; protein tyrosine kinase activator activity; receptor tyrosine kinase binding; receptor binding \\
10096233  &      8   &    32115256      &  NRG1/intron    & 7.15& (Bio)   nervous system development; regulation of protein heterodimerization activity; Notch signaling pathway; positive regulation of cell adhesion; transmembrane receptor protein tyrosine kinase activation (dimerization); neural crest cell development; cellular protein complex disassembly; wound healing; regulation of protein homodimerization activity; ventricular cardiac muscle cell differentiation; positive regulation of striated muscle cell differentiation; positive regulation of cell growth; cardiac muscle cell differentiation; cell proliferation; embryonic development; mammary gland development; anti-apoptosis; cell communication; negative regulation of secretion; negative regulation of transcription, DNA-dependent; transmembrane receptor protein tyrosine kinase signaling pathway; positive regulation of cardiac muscle cell proliferation\\

10756461  &      9   &    13185149      &  MPDZ/intron    & 7.79&  (Mol)   protein C-terminus binding; protein binding\\
	&		&		&		&	& (Bio)   interspecies interaction between organisms\\

1538677   &      10  &    72543579  &      C10orf27/intron &8.11& (Bio)   multicellular organismal development; spermatogenesis; cell differentiation\\
12415051  &      10  &    72543913  &      C10orf27/intron &8.36&\\ 
10998340  &      10  &    70383593  &      TET1/intron     &8.24&(Mol)   oxidoreductase activity, acting on single donors with incorporation of molecular oxygen, incorporation of two atoms of oxygen; structure-specific DNA binding; zinc ion binding; iron ion binding; metal ion binding; oxidoreductase activity\\
	&		&		&		&	& (Bio)    inner cell mass cell differentiation; regulation of transcription, DNA-dependent; stem cell maintenance; chromatin modification\\
\end{tabular}
\end{center}
\end{table}

\begin{table}[H]
\scriptsize
\legend{TABLE S3: (bis)} 
\begin{center}
\begin{tabular}{c|c|c|c|c|p{12cm}}				
SNP   &          CHR  &   BP    &          Gene     &       |z|-score	&	Gene function\\
\hline
2403221  &       11  &    9852475   &      SBF2/intron   &  8.33& (Mol)   protein binding; protein homodimerization activity; phosphatase regulator activity; phosphoinositide binding; phosphatase binding\\
	&		&		&		&	& (Bio)   myelination; protein tetramerization\\
4910295  &       11  &    11311743  &      GALNTL4/intron&  7.27& (Mol)   transferase activity, transferring glycosyl groups; polypeptide N-acetylgalactosaminyltransferase activity; sugar binding \\

11066776 &       12  &    114264827 &      RBM19/intron  &  7.04& (Mol)   RNA binding; nucleotide binding\\
	&		&		&		&	& (Bio)   positive regulation of embryonic development; multicellular organismal development\\

9543476  &       13  &    74425228   &     KLF12/intron &   7.16&  (Mol)   DNA binding; zinc ion binding; metal ion binding; transcription corepressor activity; transcription factor activity\\
	&		&		&		&	& (Bio)   regulation of transcription from RNA polymerase II promoter; positive regulation of transcription from RNA polymerase II promoter; negative regulation of transcription from RNA polymerase II promoter\\

1760907  &       14  &    20844859    &    TEP1/intron &    7.15& (Mol)   telomerase activity; RNA binding; nucleotide binding; ATP binding\\
	&		&		&		&	& (Bio)   telomere maintenance via recombination\\

6063071 &        20  &    45737763   &     EYA2/intron  &   7.09& (Mol)   protein binding; hydrolase activity; magnesium ion binding; protein tyrosine phosphatase activity\\
	&		&		&		&	& (Bio)   istone dephosphorylation; striated muscle development; regulation of transcription, DNA-dependent; apoptosis; multicellular organismal development; mesodermal cell fate specification; chromatin modification; DNA repair\\

2294352    &     22  &    40827319   &     MKL1/intron  &   7.96& (Mol)   actin monomer binding; leucine zipper domain binding; protein binding; nucleic acid binding; transcription coactivator activity\\
3827382    &     22  &    40881403   &     MKL1/intron  &   7.73& (Bio)   positive regulation of transcription, DNA-dependent; anti-apoptosis; smooth muscle cell differentiation; positive regulation of transcription from RNA polymerase II promoter\\
6001912    &     22  &    40828361   &     MKL1/intron  &   7.26&\\
6001913    &     22  &    40836753   &     MKL1/intron  &   7.51&\\
17002034   &     22  &    40996367   &     MKL1/intron  &   8.01&\\

5917471    &     X   &    37652518   &     CYBB/intron  &   7.04& (Mol)    protein binding; FAD binding; electron carrier activity; protein heterodimerization activity; metal ion binding; superoxide-generating NADPH oxidase activity; heme binding; voltage-gated ion channel activity; oxidoreductase activity\\
	&		&		&		&	& (Bio)   respiratory burst; superoxide metabolic process; innate immune response; ion transport; inflammatory response; superoxide release; hydrogen peroxide biosynthetic process\\
5944708    &     X   &    25000842   &     POLA1/intron &   7.63& (Mol)   DNA primase activity; metal ion binding; nucleotide binding; DNA-directed DNA polymerase activity; nucleotidyltransferase activity; transferase activity; protein binding; DNA binding; protein heterodimerization activity; purine nucleotide binding; double-stranded DNA binding; nucleoside binding; chromatin binding; pyrimidine nucleotide binding\\
	&		&		&		&	& (Bio)   DNA replication initiation; M/G1 transition of mitotic cell cycle; interspecies interaction between organisms; DNA replication, synthesis of RNA primer; DNA strand elongation during DNA replication; leading strand elongation; DNA repair; lagging strand elongation; double-strand break repair via nonhomologous end joining; telomere maintenance via semi-conservative replication; G1/S-specific transcription in mitotic cell cycle; cell proliferation; DNA synthesis during DNA repair; telomere maintenance via recombination; nucleobase, nucleoside, nucleotide and nucleic acid metabolic process; mitotic cell cycle; DNA replication checkpoint; S phase of mitotic cell cycle; DNA replication; telomere maintenance; G1/S transition of mitotic cell cycle\\
6643647    &     X   &    153086372  &     PDZD4/intron &   7.95& --\\
 
\end{tabular}
\end{center}
\end{table}
\end{landscape}

%% file: suppmat1.tex
\section*{Gibbs Sampling algorithm for the LFMM}
\small

\subsection{Prior Distribution}

Let $D$ is the number of environmental variables. $I_{i,\ell}$ is the indicator variable equal to 0 if the data are missing and 1 otherwise.
${\rm N}(\mu,\Sigma)$ is the normal distribution of mean $\mu$ and of covariance matrix $\Sigma$. $\Gamma^{-1}(a,b)$ is the inverse-gamma distribution of shape $a$ and of rate $b$ (ie of scale $\frac{1}{b}$).

The prior distributions on the LFMM parameters are given by:
\begin{equation}
\textit{for all } i,\ell ~~~ G_{i,\ell} | U_{i},V_{\ell},\beta_{\ell},\mu_{\ell},\sigma^2 
\sim {\rm N} (X_{i}\beta_{\ell} + \mu_{\ell} + U_{i}^{T}V_{\ell},\sigma^2)^{I_{i,\ell}}
\end{equation}

\begin{equation}
\textit{for all } i ~~~ U_{i} | \sigma_{U}^{2} \sim {\rm N} (0,\sigma_{U}^{2}I_{K})
\end{equation}

\begin{equation}
\textit{for all } \ell ~~~ V_{\ell} | \sigma_{V}^2 \sim {\rm N}(0,\sigma_{V}^{2}I_{K})
\end{equation}

\begin{equation}
\textit{for all } \ell,d ~~~ \beta_{d,\ell} | \sigma_{\beta(d)}^{2} \sim {\rm N}(0,\sigma_{\beta(d)}^{2})
\end{equation}

\begin{equation}
\textit{for all } \ell ~~~ \mu_{\ell} | \sigma_{\mu}^{2} \sim {\rm N}(0,\sigma_{\mu}^{2})
\end{equation}

$\sigma_{V}^{2} = 1$ and $\sigma^2$ is updated at each iteration by the current residual variance.\\

$\textit{for all } d ~~~ \sigma_{\beta(d)}^2$, $\sigma_{\mu}^2$ and $\sigma_U^2$ follow an inverse-gamma distribution $\Gamma^{-1}(\eta,\eta)$ where $\eta = 10^{3}$. 


\subsection{Conditional Distribution}

The LFM Model is a hierarchical model which conditional distributions can be described as follows:

\begin{equation}
p(\sigma_U^2 | U,\eta) = \Gamma^{-1}(\eta + \frac{NK}{2},\frac{1}{2}\sum_i U_{i}^{T}U_{i} + \eta)
\end{equation}

\begin{equation}
p(\sigma_{\beta(d)}^2 | \beta,\eta) = \Gamma^{-1}(\eta + \frac{M}{2},\frac{1}{2}\sum_l \beta_{d,\ell}^{2} + \eta)
\end{equation}

\begin{equation}
p(\sigma_{\mu}^2 | \mu,\eta) = \Gamma^{-1}(\eta + \frac{M}{2},\frac{1}{2}\sum_\ell \mu_{\ell}^{2} + \eta)
\end{equation}

\begin{equation}
p(U_{i} | G,V,\beta,\mu,\sigma_U^2,\sigma^2) = {\rm N}(\mu_{U}^{i},{\Delta_{U}^{i}}^{-1})
\end{equation}
where
\begin{equation}
\Delta_{U}^{i} = {\sigma_{U}^2}^{-1} I_{K} + {\sigma^2}^{-1} \sum_\ell V_{\ell} V_{\ell}^{T}
\textit{ and }
\mu_{U}^{i} = {\sigma^2}^{-1} (\Delta_{U}^{i})^{-1} \sum_\ell (G_{i,\ell} - X_{i}\beta_{\ell} - \mu_{\ell}) V_{\ell}
\end{equation}
 
\begin{equation}
p(V_{\ell} |G,U,\beta,\mu,\alpha_G) = {\rm N}(\mu_{V}^{\ell},{\Delta_{V}^{\ell}}^{-1})
\end{equation}
where
\begin{equation}
\Delta_{V}^{\ell} = {\sigma_{V}^2}^{-1} I_{K} + {\sigma^2}^{-1} \sum_i U_{i} U_{i}^{T}
\textit{ and }
\mu_{V}^{\ell} = {\sigma^2}^{-1}(\Delta_{V}^{\ell})^{-1} \sum_i (G_{i,\ell} - X_{i}\beta_{\ell} - \mu_{\ell}) U_{i}
\end{equation}

\begin{equation}
p(\beta_{\ell} |G,U,V,\mu,\sigma_\beta(1)^2,\ldots,\sigma_\beta(d)^2,\sigma^2) = {\rm N}(\mu_{\beta}^{\ell},{\Delta_{\beta}^{\ell}}^{-1})
\end{equation}
where
\begin{equation}
\Delta_{\beta}^{\ell} = diag({\sigma_\beta(1)^2}^{-1},\ldots,{\sigma_\beta(d)^2}^{-1}) + {\sigma^2}^{-1} \sum_i X_{i}^{T} X_{i}
\textit{ and }
\mu_{\beta}^{\ell} = {\sigma^2}^{-1} (\Delta_{\beta}^{\ell})^{-1} \sum_i (G_{i,\ell} - U_{i}^{T}V_{\ell} - \mu_{\ell}) X_{i}^{T}
\end{equation}

\begin{equation}
p(\mu_{\ell} |G,U,V,\beta,\sigma_\mu^2,\sigma^2) = {\rm N}(\mu_{\mu}^{\ell},{\Delta_{\mu}^{\ell}}^{-1})
\end{equation}
where
\begin{equation}
\Delta_{\mu}^{\ell} = {\sigma_\mu^2}^{-1} + {\sigma^2}^{-1} N
\textit{ and }
\mu_{\mu}^{\ell} = {\sigma^2}^{-1} (\Delta_{\mu}^{\ell})^{-1} \sum_{i} (G_{i,\ell} - U_{i}^{T}V_{\ell} - X_{i}\beta_{\ell}) 
\end{equation}

\subsection{Main algorithm}
$nIter$ is the number of iterations and $burn$ is the number of iterations for burning.\\
\fbox{
\begin{minipage}{1.0\textwidth} 
\begin{enumerate}
\item Initialize model parameters
$$ U = 0_{K,N}$$
$$ V = 0_{K,M}$$
$$ \beta = 0_{M,D}$$
$$ \mu = 0_{M,1}$$
\item For $n=1\ldots nIter$
\begin{itemize}
\item input missing values at locus $\ell$ for individu $i$, 
$$G_{i,\ell} \leftarrow {{U_i}^{(n-1)}}^{T}V_\ell^{(n-1)}+X_i^{(n-1)}\beta_\ell^{(n-1)}$$
\item update the residual variance
$$ {\sigma^2}^{(n)} = \textit{var}(G-{{U}^{(n-1)}}^{T}V^{(n-1)}-X^{(n-1)}\beta^{(n-1)})$$
\item sample the hyperparameters
$${\sigma_U^2}^{(n)} \sim p(\sigma_U^2 | U^{(n-1)},\eta)$$
$${\sigma_\beta^2}^{(n)} \sim p(\sigma_\beta^2 | \beta^{(n-1)},\eta)$$
$${\sigma_\mu^2}^{(n)} \sim p(\sigma_\mu^2 | \mu^{(n-1)},\eta)$$
\item for each locus $\ell$, sample
$$ \mu_{\ell}^{(n)} \sim p(\mu_{\ell} | U^{(n-1)},V^{(n-1)},\beta^{(n-1)},{\sigma_\mu^2}^{(n)},{\sigma^2}^{(n)})$$
$$ \beta_{\ell}^{(n)} \sim p(\beta_{\ell} | U^{(n-1)},V^{(n-1)},\mu^{(n)},{\sigma_\beta(1)^2}^{(n)},\ldots,{\sigma_\beta(d)^2}^{(n)},{\sigma^2}^{(n)})$$
\item for each individu $i$, sample
$$ U_{i}^{(n)} \sim p(U_{i} | \mu^{(n)},V^{(n-1)},\beta^{(n)},{\sigma_U^2}^{(n)},{\sigma^2}^{(n)})$$
\item for each locus $l$, sample
$$ V_{\ell}^{(n)} \sim p(V_{\ell} | \mu^{(n)},U^{(n)},\beta^{(n)},{\sigma^2}^{(n)})$$
\end{itemize}
\item compute the parameters
$$ U = \textit{mean}(U^{(burn+1)},\ldots,U^{(nIter)})$$
$$ V = \textit{mean}(V^{(burn+1)},\ldots,V^{(nIter)})$$
$$ \beta = \textit{mean}(\beta^{(burn+1)},\ldots,\beta^{(nIter)})$$
$$ \mu = \textit{mean}(\mu^{(burn+1)},\ldots,\mu^{(nIter)})$$
$$ Z = \textit{mean}(\beta^{(burn+1)},\ldots,\beta^{(nIter)})/\textit{var}(\beta^{(burn+1)},\ldots,\beta^{(nIter)})^{\frac{1}{2}}$$
\end{enumerate}
\end{minipage}
}